\documentclass[USenglish,twocolumn]{article}

\ifx\directlua\undefined\ifx\XeTeXcharclass\undefined
  \usepackage[utf8]{inputenc}                           
  \else\RequirePackage[no-math]{fontspec}[2017/03/31]\fi 
  \else\RequirePackage[no-math]{fontspec}[2017/03/31]\fi 
\usepackage[sort&compress,square,numbers]{natbib}
\usepackage[big,online]{dgruyter}

\usepackage{ulem,color}
\newcommand{\dlt}[1]{}
\newcommand{\del}[1]{}
\newcommand{\new}[1]{#1}

\theoremstyle{dgthm}

\theoremstyle{dgdef}

\begin{document}

\articletype{Research Article}

\author[1]{Rostislav Řepa}
\author[2]{Michal Horák}
\author[3]{Tomáš Šikola}
\author*[4]{Vlastimil Křápek} 
\affil[1]{Brno University of Technology, Brno, Czechia, rostislav.repa@vutbr.cz; 0009-0005-2483-3597}
\affil[2]{Brno University of Technology, Brno, Czechia, michal.horak2@ceitec.vutbr.cz; 0000-0001-6503-8294}
\affil[3]{Brno University of Technology, Brno, Czechia, sikola@fme.vutbr.cz; 0000-0003-4217-2276}
\affil[4]{Brno University of Technology, Brno, Czechia, krapek@vutbr.cz; 0000-0002-4047-8653}
\title{Charge reservoir as a design concept for plasmonic antennas}
\runningtitle{Charge reservoir in plasmonic antennas}
\abstract{
Plasmonic antennas exploit localized surface plasmons to shape, confine, and enhance electromagnetic fields with subwavelength resolution. The field enhancement is contributed to by various effects, such as the inherent surface localization of plasmons or the plasmonic lightning-rod effect. Inspired by nanofocusing observed for propagating plasmons, we test the hypothesis that plasmonic antennas with a large cross-section represent a large charge reservoir, enabling large induced charge and field enhancement. Our study reveals that a large charge reservoir is accompanied by large radiative losses, which are the dominant factor, resulting in a low field enhancement. \dlt{We then isolate the net charge reservoir effect adjusted for modified radiative losses, and show that it positively correlates with the field enhancement.}
}

\maketitle

\section{Introduction} 

Plasmonic antennas~\cite{Novotny2011} (PAs) are metallic nanostructures allowing for subwavelength control of light. They are particularly attractive for their ability to locally reshape the driving field (often a plane electromagnetic wave), including its significant confinement and enhancement. The locally enhanced field is exploited in numerous applications, including plasmon-enhanced spectroscopy~\cite{Kinkhabwala2009,Pfeiffer2010,Wang2020,Han2022,https://doi.org/10.1002/smtd.202100376}, energy harvesting~\cite{BORISKINA2013375}, strong light-matter coupling~\cite{doi:10.1021/nl104352j,Bitton2020}, or sensing~\cite{annurev:/content/journals/10.1146/annurev.physchem.58.032806.104607,Riley2023,Mejia-Salazar2018}. The mechanism of the field enhancement consists in the excitation of localized surface plasmons (LSPs), quasiparticles formed by oscillations of the free electron gas in a metal and the induced electromagnetic field. 

The design of PAs with desired properties and their optimization for specific applications exploit various concepts, including plasmonic circuit models~\cite{PhysRevLett.95.095504,Zhu:14,Benz:15,Hughes2016}, transformation optics~\cite{doi:10.1021/nl303377g,doi:10.1021/acs.jpcc.2c04828}, hybridization models~\cite{doi:10.1126/science.1089171,krapek_independent}, or Babinet's principle~\cite{PhysRevB.76.033407,doi:10.1021/nl402269h,Horak2019,10.1063/5.0065724}. A particularly inspiring concept is nanofocusing~\cite{10.1063/1.372414,PhysRevLett.93.137404,doi:10.1021/nl071340m,PhysRevLett.105.116804,Schnell2011,Choo2012,Gramotnev2014,9172141}, originally formulated for {\it propagating} surface plasmon polaritons: When they propagate in the plasmonic structure with a decreasing cross-section (e.g., a tapered waveguide or a metallic tip for near-field microscopy), the electromagnetic energy is concentrated. The role of the nanofocusing for LSPR supported by PAs is less clear. The effect of the narrow part is well known. The so-called plasmonic lightning-rod effect~\cite{Garcia-Etxarri:12,Shi:17,Jie_Yang,Rogobete:07,Urbieta2018,krapek_PLRE} states that the induced electric field is particularly strong at the sharp parts of PAs \new{(differences between the electrostatic and plasmonic lightning-rod effects are discussed in Ref.~\cite{krapek_PLRE})}. However, the broad part of the PA is only rarely considered, although it is an essential part of nanofocusing. Several exceptions are represented by a comparison of a bowtie PA and a nanorod dimer PA, showing a stronger field enhancement of the latter~\cite{Biagioni_2012}, a study focusing on the effect of a vortex angle in a triangular PA~\cite{10.1063/1.1602956}, \new{a comparison of the split ring resonator and crescent resonator showing a stronger field of the latter~\cite{Chen:17},} and a comparison of the field enhancement achieved through nanofocusing and with a PA~\cite{Li:25}. However, a systematic study of nanofocusing in PAs, including the analysis of the broad part, is missing. 

We propose a hypothesis that the broad part can act as a reservoir of the charge for plasmonic oscillations; when driven by an external excitation, the broad part provides electrons that can be concentrated at the narrow part, forming a large induced charge accompanied by a large electric near field. In our paper, this hypothesis is tested and invalidated.

\section{Charge reservoir}

The central question addressed in our work reads: Can we expect a stronger response of plasmonic antennas if they contain more electrons, under otherwise identical circumstances? In this section, we will explain the motivation for this question, refine it, and design a study to find the answer. 

\begin{figure*}
\includegraphics[width=2\columnwidth]{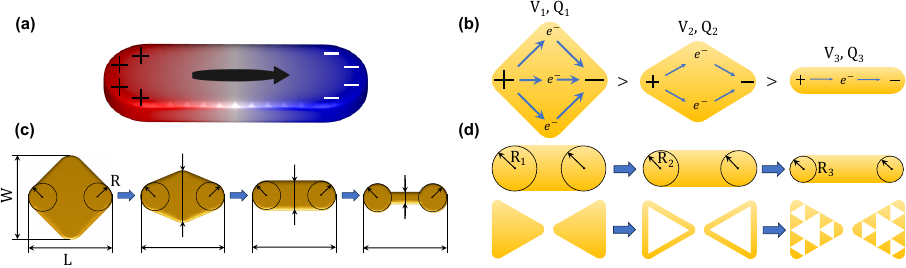}
\caption{
(a) A scheme of a dipole LSPR supported by a plasmonic nanorod. The current is represented by the black arrow, thickened in the central part of the PA to mark the antinode of the current wave. The antinodes of the induced charge at the peripheral parts of the PA are labeled by $+$ and $-$ symbols. (b) Conceptual explanation of the charge reservoir. From left to right, the cross-section of PAs decreases, resulting in expectedly lower magnitudes of induced charge (represented by decreased size of $+$ and $-$ symbols). (c) The design of the study. From left to right, a diamond PA is transformed through a tapered-diamond PA into a nanorod PA and then to a dumbbell PA by a sequential thinning of the central part, accompanied by a reduction in the charge reservoir. The black circles at the peripheral part of all PAs demonstrate that the radius of curvature is preserved for all PAs. (d) Two sets of PAs not suitable for studying the charge reservoir effect. Top: The nanorods of different widths also differ in the radius of curvature. Bottom: Three bowtie PAs (full, contour, and fractal) differ in the charge oscillation patterns.
\label{fig1}
}
\end{figure*}

We will consider only planar PAs that can be fabricated by the lithography of a thin metallic film. The amount of electrons within the PA will be controlled only by its shape (i.e., the geometrical cross-section); alteration of the PA material is not considered. We will focus on a dipole LSPR, which can be visualized as an oscillating electric dipole, with the antinodes of the oscillating charge at peripheral parts of the PA and the current antinode in the central part, as schematically shown in Fig.~\ref{fig1}(a). 

Fig.~\ref{fig1}(b) explains the idea of the charge reservoir. Three PAs differing in the cross-section and identical otherwise are subject to the same excitation (a plane electromagnetic wave \new{with the polarization} parallel \new{to}\dlt{with} the excited dipole). In the PAs with a larger cross-section, the driving field shall affect a large charge, and it is plausible to assume that the charge induced at the peripheral parts would be larger in magnitude, as schematically indicated in Fig.~\ref{fig1}(b). In other words, the volume of the PA shall act as a {\it reservoir} of the charge that can contribute to LSPR. We will investigate whether this effect, labeled as the {\it charge reservoir \new{effect}}, really exists, is observable, and can be exploited in specific applications of PAs. To this end, we will carry out a comparative study of PAs with different charge reservoirs.

Multiple effects affect the strength of the response of PAs. To isolate the role of the charge reservoir, it is essential to compare the PAs differing in the charge reservoir under otherwise identical conditions. This includes identical material composition and excitation field. We will also aim to keep the LSPR energy constant, either approximately by fixing the length of the PA, or exactly by adjusting the length to the specific energy. The induced field is enhanced near the curved features of the PAs (so-called lightning-rod effect)~\cite{Garcia-Etxarri:12,krapek_PLRE}. We will thus keep constant the geometrical curvature of those parts of PAs where the field is evaluated. Finally, we will require that all PAs support the lowest-energy LSPR of the same plain-dipole topology. Thus, concepts like contour~\cite{Sederberg:11,doi:10.1021/jp408610q} or fractal~\cite{Sederberg:11b} PAs with the inner part of the PA removed will not be considered in our study. Although they provide a clear way to modify the charge reservoir, the overall performance would be influenced by additional effects, among which the coupling might be particularly important~\cite{doi:10.1021/jp408610q}. 

In our study, we will take a simple plasmonic rod with rounded tips as the central shape. Then, we will either widen the central part by gradually morphing the rod to a diamond or narrow the central part by morphing the rod to a dumbbell. This set of PAs, schematically represented in Fig.~\ref{fig1}(c), will be used to investigate the charge reservoir. Fig.~\ref{fig1}(d) then shows two sets of PAs which exhibit a variation of the charge reservoir combined with a variation of additional factors, which are unsuitable for isolating the role of the charge reservoir.

\section{Methods}

Our methodology combines experimental electron energy loss spectroscopy (EELS) with electromagnetic simulations. EELS provides qualitative insight into the near field of PAs and verifies the simulations, which provide additional quantities such as a full quantitative characterization of the electric field.

\subsection{Fabrication of plasmonic antennas}
PAs were fabricated by the focused ion beam (FIB) lithography~\cite{Horak2018}. A gold layer with a thickness of 30 nm was deposited by ion beam sputtering using an optimized deposition rate of 0.1 nm/s~\cite{doi:10.1021/acsomega.4c06598} onto the standard 30-nm-thick silicon nitride membranes for transmission electron microscopy (Agar Scientific). FIB milling was performed using the dual-beam FIB/SEM microscope Tescan LYRA3 with the gallium ion beam energy set to 30 keV and the beam current set to 1 pA (lowest available). PAs were situated in the centre of a metal-free square with the dimensions of $2\times 2\ \mathrm{\text{\textmu}m}^2$, which is sufficient to prevent their interactions with the surrounding metallic frame and the neighbouring PAs.

\subsection{Electron energy loss spectroscopy}  
EELS measurements were performed using a FEI Titan S/TEM equipped with a GIF Quantum spectrometer. The microscope was operated in scanning monochromated mode at 120 keV, with convergence and collection semi-angles set to 10 and 11.4 mrad, respectively, and spectrometer dispersion set to 0.01 eV/pixel. The probe current was adjusted to around 100 pA, and the full width at half maximum (FWHM) of the zero-loss peak was found to be in the range of 0.10 to 0.12 eV. To use the full intensity range of the CCD camera in the spectrometer and avoid overexposure, the acquisition time of every spectrum was set to 0.5 ms. Before performing the EELS measurements, the sample was plasma-cleaned in argon–oxygen plasma for 20 seconds to minimize carbon contamination. 

For each PA, we recorded a spectrum image with the spatial dimensions set to $500\times500\ \mathrm{nm}^2$ and the pixel size of 2 nm. The raw spectra were averaged over a certain region of interest to reduce the noise. Next, the spectra were normalized to 
the electron counts of the zero-loss peak (ZLP) (integrated over the energy window from $-1$~eV to $+1$~eV) to obtain a quantity proportional to the loss probability per channel, which was further divided by the energy interval of the channel (0.01 eV) to obtain the usual loss probability density (in units 1/eV) (the so-called loss function). Finally, the experimentally determined background and the ZLP (measured at the bare membrane) were subtracted to isolate the contribution of LSPR. For experimental loss function maps, we first integrated the spectrum image over the energy range of 0.1 eV around the resonance energy of an LSPR to obtain a signal map (with reduced noise), which was further divided pixel-wise by the map of the integral intensity of the zero-loss peak. In this case, the zero-loss peak and background were not subtracted. 

\subsection{Electromagnetic simulations} 
The electromagnetic simulations were performed using the MNPBEM toolbox~\cite{HOHENESTER2012370,HOHENESTER20141177,WAXENEGGER2015138} for Matlab based on the boundary element method (BEM)~\cite{PhysRevB.65.115418}. The dielectric function of gold was taken from~\cite{PhysRevB.6.4370}. In all simulations, the membrane was neglected, and the refractive index of the surrounding medium was set to 1. \new{Further details, including the dimensions of all simulated PAs, are provided in Supporting Information, Sec.~S1. The effect of the neglected membrane is discussed in Supporting Information, Sec.~S2.}


\section{Results}
In this section, we first validate the simulations by comparing their predictions to experimental data. We then analyze the charge reservoir effect, demonstrating that PAs with a larger charge reservoir generally exhibit a weaker response due to their increased radiative losses. Finally, we discuss the relation between the charge reservoir and specific response functions for various parameters of PAs.

\subsection{Experimental validation of simulations}
\begin{figure}
\includegraphics[width=\columnwidth]{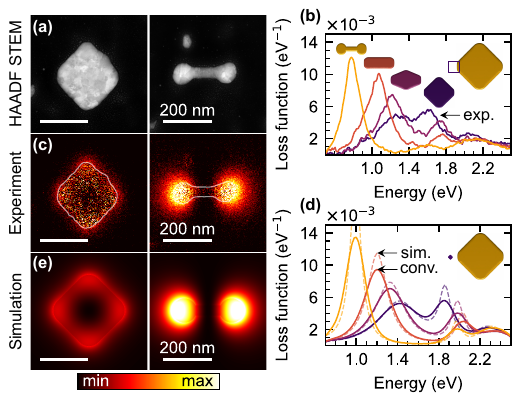}
\caption{
EELS of a set of PAs with a length $L=300$~nm, a radius $R=50$~nm, and widths $W$ between 300~nm and 40~nm. (a) HAADF STEM image of two PAs with the widths $W=300$~nm and 40~nm. (b) Experimental loss function spectra for PAs with widths $W$ of 40~nm (golden line, dumbbell shape), 100~nm (salmon line, rod shape), 183~nm (maroon line, tapered diamond shape), and 300~nm (indigo line, symmetric diamond shape). The insets above the spectra also indicate the correspondence between the spectra and the shapes. The spectra were averaged over the electron beam positions corresponding to a square next to the left edge of the PA, as schematically indicated in the right inset. (c) Experimental loss function maps recorded for PAs from panel a for the loss energy corresponding to the dipole LSPR (represented by a peak in the corresponding spectra in panel b between 1.0 and 1.4~eV). (d) Theoretical loss function spectra analogous to panel b. The electron beam position, 20~nm from the left edge of PAs,  is indicated in the right inset. The dashed lines represent raw simulated spectra, the solid lines show the spectra convolved with a Gaussian with a full width at half maximum (FWHM) of 0.15~eV representing the instrumental broadening of our EELS setup. (e) Theoretical loss function maps analogous to panel c.
\label{fig2}
}
\end{figure}

We focused on a set of planar PAs [diamonds, a rod, dumbbells as shown in Fig.~\ref{fig1}(c)] with the identical length $L=300$~nm, radius $R=50$~nm, and thickness of 30~nm, while the width $W$ of PAs was varied from 300~nm to 40~nm, with $W>100$~nm corresponding to the (tapered) diamond shape, $W=100$~nm to the rod shape, and $W<100$~nm to the dumbbell shape. The PAs were fabricated and analyzed experimentally by EELS. The EELS was also theoretically modelled. The results of this analysis are summarized in Fig.~\ref{fig2}.

Fig.~\ref{fig2}(a) shows a high-angle annular dark-field (HAADF) STEM image of two PAs of the set with $W=300$~nm (a symmetric diamond) and $W=40$~nm (a dumbbell). The experimental loss function is displayed as a spectrum for a specific position of the electron beam in Fig.~\ref{fig2}(b) and as a map for a specific value of the energy loss (corresponding to the dipole LSPR) in Fig.~\ref{fig2}(c). The loss function spectra in Fig.~\ref{fig2}(b) are shown for a symmetric diamond ($W=300$~nm), a tapered diamond ($W=183$~nm), a rod ($W=100$~nm), and a dumbbell ($W=40$~nm), as indicated by insets above each spectrum. To suppress experimental noise, the spectra were averaged over a range of electron beam positions spanning a square next to the left edge of the PA, as schematically indicated in the right inset of Fig.~\ref{fig2}(b). In the following, we focus on the lowest-energy peak of the spectra (with the energy ranging from 1.0~eV to 1.4~eV). The maps of the loss probability at the energy of this peak are shown in Fig.~\ref{fig2}(c), and assign this peak to a dipole LSPR~\cite{krapek_independent}. The calculated spectra and maps shown in Fig.~\ref{fig2}(d,e) closely match the experimental data, verifying the correctness and accuracy of the simulations and allowing us to use the simulations to retrieve quantities inaccessible experimentally, in particular, the induced electric field. We have convolved the calculated EEL spectra with a Gaussian (FWHM of 0.15~eV) to account for the instrumental broadening. We note that the dimensions of the fabricated PAs are slightly smaller (by about 10~\%) than the nominal values considered in simulations.

\new{The second peak in the loss function spectra [Fig.~\ref{fig2}(b,d)] with the energy ranging from 1.7~eV to 2.1~eV can be assigned to a longitudinal quadrupole LSPR~\cite{krapek_independent,Kejik:20}. This LSPR is dark. It does not couple to the plane-wave excitation, nor does it decay radiatively. It can be excited by the electron beam, but in simulations, it is well spectrally separated from the dipole mode. Therefore, it will not be considered in further analysis.}

\subsection{Radiative losses and \new{total plasmon charge}\dlt{net charge reservoir effect}}

\begin{figure}
\includegraphics[width=\columnwidth]{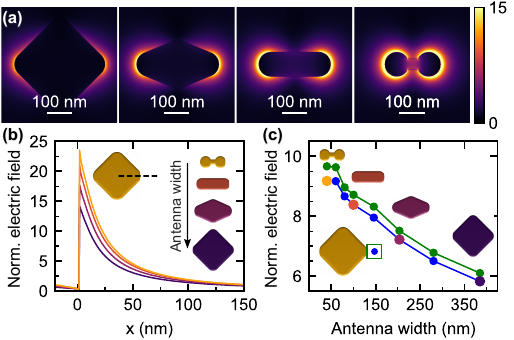}
\caption{
\new{Calculated m}\dlt{M}agnitude of the induced electric field of PAs illuminated with a plane wave at the energy of the dipole LSPR (1.2~eV). (a) A planar cross-section in the middle of the PA height for PAs with the widths of 385~nm (\new{indigo line,} symmetric diamond \new{shape}), 204~nm (\new{maroon line,} tapered diamond \new{shape}), 100~nm (\new{salmon line,} rod \new{shape}), and 40~nm (\new{golden line,} dumbbell \new{shape}). (b) A linear cross-section along the long axis of PAs (indicated in the inset) for the same PAs as in panel a. The distance from the PA edge is labeled $x$. (c) The magnitude of the field at a distance of 25~nm from the PA edge (blue) or averaged over a cube with a side of 50~nm centered at a distance of 25~nm from the PA edge (green). The fields are normalized to the amplitude of the incident plane wave. The inset shows the position/area where the field is evaluated. The symbols in panel c corresponding to the PAs included in panel b are enlarged and displayed using corresponding colors.
\label{fig3}
}
\end{figure}

\begin{figure}
\includegraphics[width=\columnwidth]{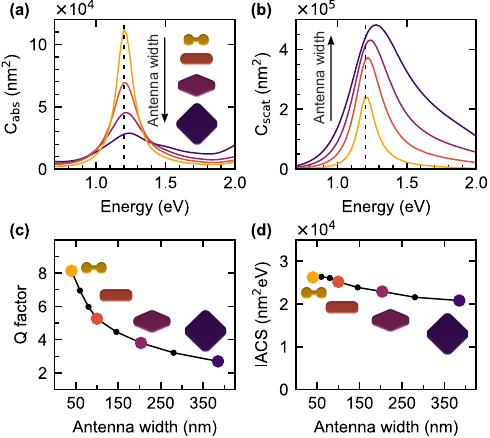}
\caption{
\dlt{O}\new{Calculated o}ptical response of a set of PAs with a length adjusted for the energy of the dipole LSPR of 1.2~eV, a radius $R=50$~nm, and widths $W$ between 385~nm and 40~nm. (a) Absorption cross-section for PAs with widths $W$ of 40~nm (golden line, dumbbell shape), 100~nm (salmon line, rod shape), 204~nm (maroon line, tapered diamond shape), and 385~nm (indigo line, symmetric diamond shape). The insets on the right side of the spectra also indicate the correspondence between the spectra and the shapes. (b) Scattering cross-section for the same set of PAs. (c) Q factor. (d) Integral absorption cross-section (IACS). The symbols in panels c and d corresponding to the PAs included in panels a and b are enlarged and displayed using corresponding colors.
\label{fig4}
}
\end{figure}

The results of the previous section, displayed in Fig.~\ref{fig2}, already show that the PAs with a lower charge reservoir exhibit, rather counterintuitively, a stronger response, as can be observed, e.g., by comparing an intense loss function of the dumbbell to a weak loss function of the diamond in Fig.~\ref{fig2}(c). However, there are two factors, besides the charge reservoir, that might contribute to this observation. (i) The energy of the dipole LSPR differs among the PAs involved in the comparison. (ii) The electron beam is a localized excitation, with a characteristic decay length~\cite{RevModPhys.82.209} of 120~nm for the energy transfer of 1.2~eV and the electron energy of 120~keV utilized in our study. To exclude the effect of these factors, we have adjusted the length of the PAs so that they all exhibit the identical energy of the dipole LSPR of 1.2~eV, and we utilized excitation with a plane electromagnetic wave \new{polarized along the long axis of PAs}. The results are summarized in Figs.~\ref{fig3} and \ref{fig4}.



The induced electric field is shown in Fig.~\ref{fig3}. Planar maps in Fig.~\ref{fig3}(a) qualitatively confirm stronger fields in narrower PAs with lower charge reservoirs. This observation is further supported by linear cross-sections shown in Fig.~\ref{fig3}(b), which demonstrate that the effect affects a whole area of the enhanced near field rather homogeneously. To quantify the effect, we plot the field at a distance of 25~nm from the PA edge and the field averages over a cube with a side of 50~nm centered around the previous spot in Fig.~\ref{fig3}(c) as functions of width.

Fig.~\ref{fig4}(a) shows the calculated spectral dependence of the absorption cross-section, which features a single peak centered at the energy of 1.2~eV, corresponding to the dipole LSPR. A behaviour similar to EELS is observed: as the width of a PA increases (and so does the charge reservoir), the amplitude of the absorption peak is reduced and the peak is broadened. This fact is explained by inspecting the scattering cross-section of the PAs [Fig.~\ref{fig4}(b)], which is markedly larger for wider PAs (with a larger charge reservoir), with a consequence of large radiative losses, which act as an additional damping channel for LSPR, decreasing and broadening the response function. This is further corroborated by the Q factor [Fig.~\ref{fig4}(c)] obtained as a ratio of the central energy and the FWHM of the absorption peak (obtained by fitting a Lorentzian to the absorption cross-section), which is pronouncedly reduced for wider PAs. Similarly, enhanced radiative losses explain the reduced field in PAs with a higher charge reservoir, as observed in Fig.~\ref{fig3}. We note that the positive correlation between the size of a PA and its scattering is well known~\cite{Biagioni_2012,10.1063/1.1602956}.

\dlt{
We will now provide a procedure to define the net charge reservoir effect by compensating for the modified radiative losses. To this end, we will employ a (crude) model of an LSPR as a damped harmonic oscillator with a charge $q$, a mass $m$, a resonance frequency $\omega_0$, and a Q factor $Q$, driven by a harmonic field $E(t)=E_0\exp(-i\omega t)$. The power $P(\omega)$ absorbed by the oscillator reads
}
\new{
Optical response functions of PAs exhibit characteristic resonant spectral dependences. Increased radiative losses result in a decrease and broadening of resonance curves. It might be instructive to complement the analysis of the peak values with the analysis of spectral integrals of the response functions over a full range of the dipole LSPR. As an illustration, we will consider a crude model of an LSPR as a damped harmonic oscillator with a charge $q$, a mass $m$, a resonance frequency $\omega_0$, and a Q factor $Q$, driven by a harmonic field $E(t)=E_0\exp(-i\omega t)$. The power $P(\omega)$ absorbed by the oscillator reads
}
\begin{equation}
P(\omega)=\frac{(qE_0)^2}{2m\omega_0}Q\frac{\gamma^2\omega^2}{(\omega_0^2-\omega^2)^2+\gamma^2\omega^2},
\label{eq1}
\end{equation}
\new{
where $\gamma=\omega_0/Q$. The peak absorbed power is proportional to the Q factor, $P_\mathrm{max}\sim Q$, and the FWHM is equal to $\gamma$ and is thus inversely proportional to the Q factor. The spectral integral of the power, $\int P(\omega) \mathrm{d}\omega$, is in this crude model approximately independent of the Q factor, and thus independent of the charge reservoir. However, in more realistic models, or for different response functions, the spectral integral might depend on the charge reservoir, providing additional insight into the relation between the charge reservoir and the response of PAs. In the following, we will classify the optical response functions into three categories: positively correlated to the charge reservoir if both the peak value and the spectral integral increase with increasing charge reservoir, negatively correlated to the charge reservoir if both the peak value and the spectral integral decrease with increasing charge reservoir, and partially correlated to the charge reservoir if the peak value decreases and spectral integral increases with increasing charge reservoir. We exploit this approach in Fig.~\ref{fig4}(d), where we inspect the spectral integral of the absorption cross-section. Remarkably, the spectral integral is nearly identical for all involved PAs, just as predicted by the crude oscillator model.
}
\dlt{
where $\gamma=\omega_0/Q$. The peak absorbed power is proportional to the Q factor, $P_\mathrm{max}\sim Q$, and the FWHM is equal to $\gamma$ and is thus inversely proportional to the Q factor. The spectral integral of the power, $\int P(\omega) \mathrm{d}\omega$, is then approximately independent of the Q factor. With this in mind, we will now revisit Fig.~\ref{fig4}(a). If we increase the charge reservoir, the enhanced radiative losses will result in a reduction of both the Q factor and absorption cross-section. However, if the peak intensity of the absorption cross-section reduces {\it less} than the Q factor, we can state that it has been positively affected by a net charge reservoir effect adjusted for the enhanced scattering. In the model of Eq.~\ref{eq1}, this would correspond to increased charge $q$. Alternatively, we can compare integral powers (independent of Q factors when all other parameters are kept constant). We exploit this approach in Fig.~\ref{fig4}(d), where we inspect the integral absorption cross-section, which is nearly identical for all involved PAs. Remarkably, this suggests a negligible correlation between the net charge reservoir and the absorption of PAs, even when the increased radiative losses are compensated.
}

\begin{figure}
\begin{center}
\includegraphics[width=\columnwidth]{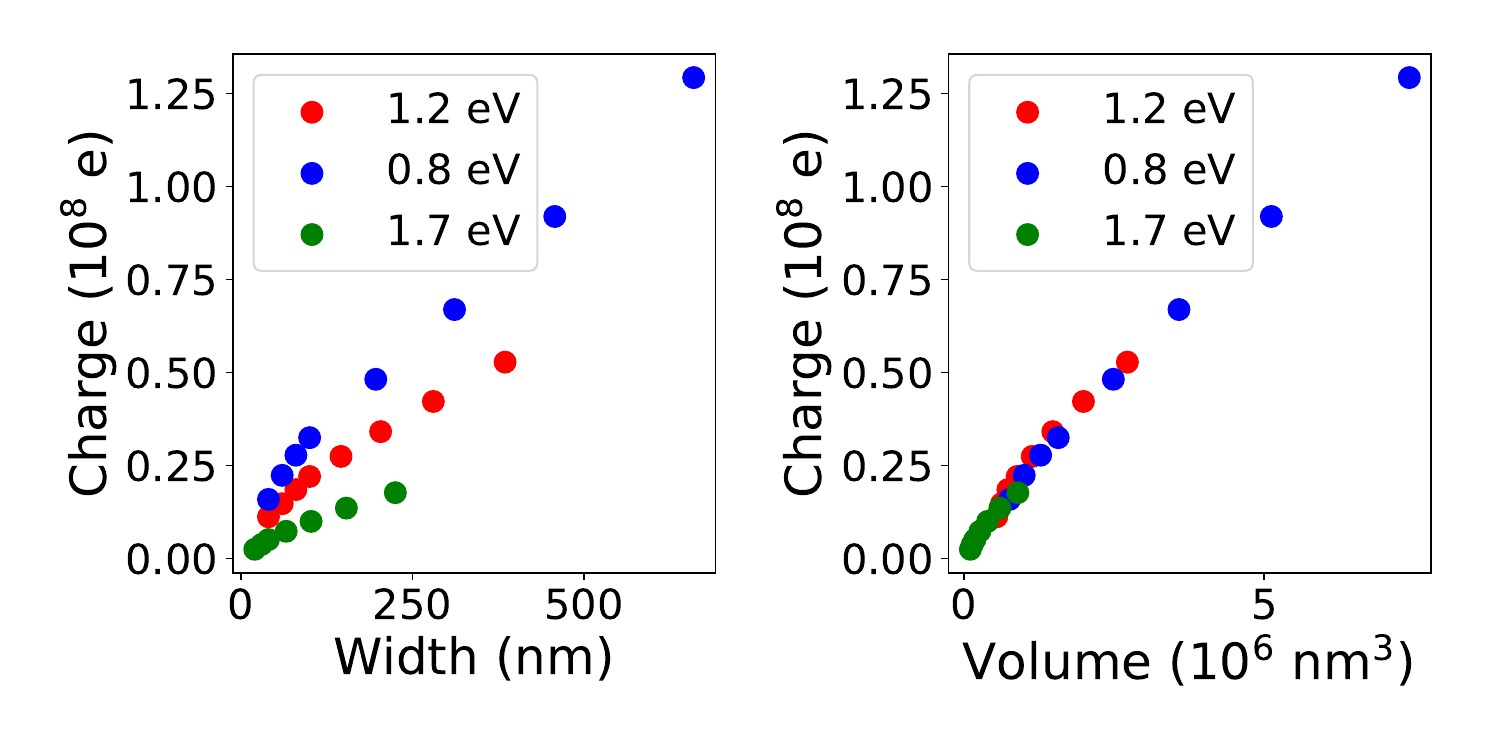}
\caption{\new{Total charge (in units of $10^8$ elementary charges) involved in plasmon oscillations according to the generalized harmonic oscillator model~\cite{Kats:11} as a function of the PA width (left) and the PA volume (right) for sets of PAs with the dipole LSPR energy fixed at 1.2~eV (red symbols), 0.8~eV (blue symbols), and 1.7~eV (green symbols).}
\label{fig5}
}
\end{center}
\end{figure}

\new{
Kats {\it et al.}~\cite{Kats:11} proposed a generalized harmonic oscillator model that correctly accounts for the radiative losses and the total charge involved in the plasmonic oscillations. We have applied the model to the calculated scattering and absorption cross-sections from Fig.~\ref{fig4} and to two other sets of PAs with the dipole LSPR energy fixed at 0.8 eV and 1.7 eV. The parameters of the model were determined by fitting the cross-section, yielding a very good agreement between the data and model. The details of the model and the fitting procedure are discussed in Supporting Information, Sec. S3. The total charge involved in plasmonic oscillations shown in Fig.~\ref{fig5} correlates with the width or the volume of PAs. Particularly prominent is the dependence on the volume, which is universal for all three sets of the PAs. In this way, the existence of the expected charge reservoir is demonstrated: The charge of the free electron participating in plasmon oscillations indeed scales with the volume of PAs. We will now investigate in more detail the relation between this charge and optical response functions.
}

\subsection{Optical response functions}

\begin{figure}
\includegraphics[width=\columnwidth]{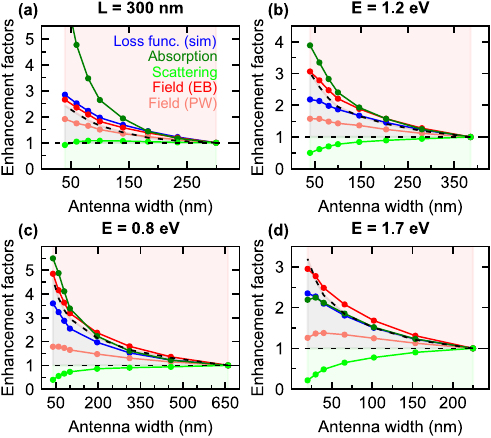}
\caption{
Enhancement factors of selected Figures of Merit (FoMs) defined as the ratio of the values for a specific PA and for a reference PA, which we set as a symmetric diamond: \dlt{Experimental and t}\new{T}heoretical peak loss probability (dark blue\dlt{ and cyan}), peak absorption cross-section (dark green), peak scattering cross-section (light green), magnitude of the electric field at a distance of 25~nm from the edge of PA for the electron beam excitation (dark red) and the plane wave excitation (light red). Four sets of PAs with different widths are compared. (a) A set of PAs with the length fixed at 300~nm. (b,c,d) Sets of PAs with the dipole LSPR energy fixed at (b) 1.2~eV, (c) 0.8~eV, and (d) 1.7~eV. The dashed black lines correspond to the enhancement factor of 1 and the enhancement factor of the Q factor, and demarcate the areas of a positive correlation with the charge reservoir (light-green shaded area), a \dlt{positive}\new{partial} correlation with the \dlt{net} charge reservoir (light-blue shaded area), and a negative correlation with the \dlt{net} charge reservoir (light-red shaded area).
\label{fig6}
}
\end{figure}

The performance of PAs in various applications is characterized by the so-called Figures of Merit (FoMs). Here we address the correlation between the charge reservoir and prominent FoMs: peak loss probability (i.e., the maximum value observed at the energy of the dipole LSPR), peak absorption and scattering cross-sections, and magnitude of the field at a specific distance from the PA edge (set to 25 nm) for both electron beam and plane wave excitations. All FoMs are retrieved from simulations\dlt{, except for the loss probability where the experimental values are also available for the set of PAs with a fixed length $L=300$~nm}. To compare the role of the charge reservoir for different FoMs, we represent them by dimensionless enhancement factors, defined as a ratio of the FoM's values for a specific PA and a reference PA (we use the symmetric diamond as the reference). The values of the enhancement factors are shown in Fig.~\ref{fig6} for four sets of PAs: with a fixed length $L=300$~nm [Fig.~\ref{fig6}(a)] and with a fixed energy of the dipole LSPR of 1.2~eV [Fig.~\ref{fig6}(b)], 0.8~eV [Fig.~\ref{fig6}(c)], and 1.7~eV [Fig.~\ref{fig6}(d)].

The analyzed FoMs can be classified into three categories: positively correlated with the charge reservoir, \dlt{positively}\new{partially} correlated with the \dlt{net} charge reservoir, and negatively correlated with the \dlt{net} charge reservoir. Those positively correlated with the charge reservoir exhibit an enhancement factor lower than 1 (the reference value for the symmetric diamond is higher than the actual value for the other PA). There is only one such FoM: the scattering cross-section. \new{Since it is demanding to calculate the spectral dependences of the induced field, we will introduce an approximate procedure to distinguish between the partial and negative correlation, assuming that the spectral integral is proportional to the peak value divided by the Q factor. Consequently, the FoMs with the enhancement factor smaller/larger than the enhancement factor of the Q factor will be classified as partially/negatively correlated to the charge reservoir, respectively. Remarkably, the induced electric field for the plane wave excitation is partially correlated to the charge reservoir. We speculate that if the radiative losses are suppressed, e.g., by employing high-Q plasmon resonances~\cite{https://doi.org/10.1002/adom.202001520}, the positive correlation between the induced electric field and the charge reservoir might be enabled.} \dlt{All the other FoMs are negatively correlated to the charge reservoir, but some of them are reduced less than the Q factor when the charge reservoir is increased. Such FoMs are classified as positively correlated with the net charge reservoir; they include the electric field associated with the plane wave excitation. The remaining FoMs are close to the interface between the positive and negative correlation to the net charge reservoir; we can thus classify them as mostly unrelated to the net charge reservoir (and negatively correlated to the charge reservoir due to radiative losses)}.

It is noteworthy that previous findings are rather general. We include in our analysis two additional sets of PAs: small ones with suppressed radiative losses (the energy of the dipole LSPR was fixed at 1.7~eV, corresponding lengths varied between 110~nm and 225~nm), and large ones which are closer to a perfect conductor regime (the energy of the dipole LSPR was fixed at 0.8~eV, the corresponding dielectric function of gold $\varepsilon=-115+11i$). While there are some variations in the values of FoMs, no systematic difference was observed.

\new{
Next, we discuss our results in the context of recent progress in the field. Numerous studies consistently report a positive correlation between radiation losses and the nanostructure volume~\cite{PhysRevLett.88.077402,KOLWAS201345,10.1063/1.5117230,HU2022100166}, in agreement with our findings. However, no study so far has addressed a direct comparison of optical response at a fixed resonance energy and fixed curvature of the antenna terminations. This distinction is crucial. In our work, we systematically vary the charge reservoir (i.e., the antenna volume) while keeping both resonance energy and end curvature constant. This enables us to isolate the role of the charge reservoir size in determining both radiative losses and field enhancement.
}

\new{
Finally, we would like to discuss the similarities and differences between the nanofocusing of propagating plasmons in tapered plasmonic waveguides (or similar structures) and the charge reservoir effect in PAs. Both cases involve tapered geometries. The taper angles are typically, but not necessarily, smaller in the case of nanofocusing. It has been demonstrated that adiabatic nanofocusing occurs in gold up to the taper angle of 35$^\circ$, and that non-adiabatic nanofocusing occurring at larger taper angles still yields a large field enhancement~\cite{10.1063/1.2963699}. A significant difference is presented by the length of the plasmonic structure. The optimum taper length for nanofocusing of around 10~$\mathrm{\upmu m}$ has been found for gold at 633~nm (1.96~eV)~\cite{10.1063/1.2963699}, while the length of PAs operating at the same energy can be estimated as 100--200~nm, depending on the precise geometry (our PAs operating at 1.7~eV have the lengths between 110~nm and 225~nm). Consequently, the tapers used for nanofocusing support a quasicontinuous spectrum of plasmons, which results in their broadband functionality~\cite{doi:10.1126/sciadv.aba4179,Ma2021,doi:10.1021/acs.jpcc.0c11541}. On the other hand, PAs support a discrete spectrum of resonances. Another significant difference is constituted by the radiative losses, which are negligible in the case of nanofocusing but play a decisive role in the case of the charge reservoir effect, as we have demonstrated in this study. 
}

\section{Conclusion}
We have tested the hypothesis that the response of the plasmonic antennas can be enhanced by their large charge reservoir. The charge reservoir \dlt{is intuitively understood as}\new{was related to} the volume of the plasmonic antenna containing the free electron gas collectively responding to the driving electromagnetic field. The large charge reservoir (typically achieved through a large cross-section of the plasmonic antenna) might then, under otherwise identical conditions, provide large induced charges, fields, and the optical response functions such as scattering and absorption cross-sections. Our analysis, which is based on a comparative study of plasmonic antennas with varied charge reservoirs but fixed curvatures and plasmon resonance energies, revealed that only the scattering cross-section is positively correlated to the charge reservoir. This, in turn, enhances the radiative losses which result in the negative correlation between the charge reservoir and the \new{peak values of the} induced fields and response functions. \dlt{We then define the net charge reservoir effect, cleared of the variations in the radiative losses, and show that it is positively correlated with the magnitude of the induced electric field.} \new{On the other hand, spectral integrals of the induced field are positively correlated to the charge reservoir.}

The charge reservoir effect is analogous to the nanofocusing observed in propagating surface plasmon polaritons. The principal difference stems from the non-radiative character of propagating plasmon polaritons, which opens the possibility for the field enhancement through nanofocusing.

We conclude that the charge reservoir effect can be exploited as the design principle for plasmonic antennas with an efficient coupling to radiative modes, but, somewhat counterintuitively, not for the large field enhancement.


\begin{funding}
We acknowledge support from the \href{http://dx.doi.org/10.13039/501100001823}{\underline{Ministry of Education}, Youth, and Sports of the Czech Republic}, projects No. CZ.02.01.01/00/22008/0004572 (QM4ST) and LM2023051 (CzechNanoLab). RŘ was supported by \href{http://dx.doi.org/10.13039/501100004585}{\underline{Brno University of Technology}} (project No.~CEITEC VUT-J-25-8860).
\end{funding}

\begin{authorcontributions}
All authors have accepted responsibility for the entire content of this manuscript and approved its submission.
\end{authorcontributions}

\begin{conflictofinterest}
Authors state no conflict of interest.
\end{conflictofinterest}

\begin{dataavailabilitystatement}
The datasets generated during and/or analyzed during the current study are available in the Zenodo repository, doi:10.5281/zenodo.17552778. 
\end{dataavailabilitystatement}

\clearpage

\twocolumn[\section*{{\LARGE Supporting Information}}]

\renewcommand{\thefigure}{S\arabic{figure}}
\renewcommand{\thesection}{S\arabic{section}}
\renewcommand{\thetable}{S\arabic{table}}
\renewcommand{\theequation}{S\arabic{equation}}

\setcounter{figure}{0}
\setcounter{section}{0}
\setcounter{table}{0}
\setcounter{equation}{0}

\section{Electromagnetic simulations} 
The electromagnetic simulations were performed using the MNPBEM toolbox~\cite{HOHENESTER2012370,HOHENESTER20141177,WAXENEGGER2015138} for MATLAB. The MNPBEM toolbox is based on the boundary element method (BEM), which was adapted to describe electron energy loss spectroscopy (EELS) by F. J. García de Abajo in collaboration with A. Howie~\cite{PhysRevB.65.115418}. In MNPBEM, Maxwell’s equations are solved in the form of surface integral equations at the interfaces between different media. These interfaces are discretized into small boundary elements, assuming that each medium is homogeneous, isotropic, and separated by sharp (abrupt) interfaces. Once the excitation scheme is specified, the BEM equations are solved for the given excitation by computing the auxiliary surface charges and currents. From these surface charges and currents, one can then calculate derived quantities such as absorption and scattering cross sections for the plane-wave excitation, the loss function spectra in EELS simulations, the induced electromagnetic fields, etc.

\begin{figure*}[ht!]
\includegraphics[width=1.5\columnwidth]{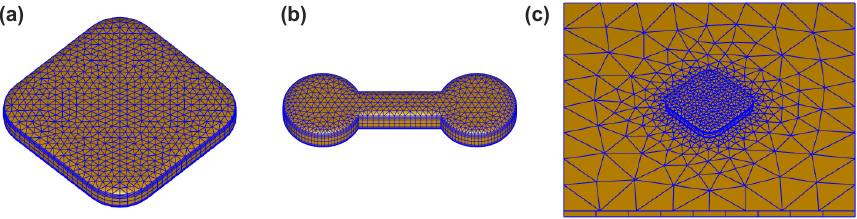}
\centering
\caption{
Examples of plasmonic antenna (PA) geometries used in the simulations, showing their discretized boundaries: (a) diamond PA, (b) dumbbell PA, and (c) diamond PA positioned on a 30-nm-thick SiN substrate.
\label{fig1SI}
}
\end{figure*}

In our study, we consider planar gold plasmonic antennas (PAs), which are modeled in the toolbox using the {\it comparticle} function. Two examples of these particle models with discretized boundaries are shown in Fig.~\ref{fig1SI}(a,b). The dielectric function of gold was taken from Ref.~\cite{PhysRevB.6.4370}, and the refractive index of the surrounding medium was set to 1. In all simulations, the BEM solver was configured to include retardation effects, i.e., to solve the full Maxwell equations, since our PAs do not satisfy the conditions required for applying the quasistatic approximation. Two types of incident excitation fields were used: an electron beam and a plane electromagnetic wave.

\begin{table*}[ht!]
  \centering
  \caption{Geometrical parameters of the simulated sets of plasmonic antennas presented in the main paper.}
  \label{tab1}
  \begin{tabular}{@{\hspace{6pt}}ccccccccc@{\hspace{6pt}}}
    \toprule
    \multicolumn{9}{c}{\textbf{Identical lengths (300~nm)}} \\ 
    \midrule
    Width (nm) & 300 & 234 & 183 & 140 & 100 & 80 & 60 & 40 \\ 
    \midrule
    \multicolumn{9}{c}{\textbf{Identical energies (1.2~eV)}} \\ 
    \midrule
    Width (nm)  & 385 & 280 & 204 & 146 & 100 & 80 & 60 & 40 \\
    Length (nm) & 385 & 370 & 350 & 330 & 305 & 283 & 261 & 236 \\
    \midrule    
    \multicolumn{9}{c}{\textbf{Identical energies (0.8~eV)}} \\ 
    \midrule
    Width (nm)  & 660 & 457 & 311 & 196 & 100 & 80 & 60 & 40 \\
    Length (nm) & 660 & 635 & 610 & 585 & 525 & 500 & 460 & 400 \\
    \midrule
    \multicolumn{9}{c}{\textbf{Identical energies (1.7~eV)}} \\ 
    \midrule
    Width (nm)  & 225 & 154 & 102 & 66 & 40 & 30 & 20 &  \\
    Length (nm) & 225 & 210 & 190 & 170 & 147 & 130 & 110 &  \\
    \bottomrule
  \end{tabular}
\end{table*}

For the electron beam (EELS) simulations, the beam parameters were identical to those used in the experimental setup. Specifically, the incident beam energy was set to 120~keV, and its position was 20~nm from the left edge of each individual PA. The loss function spectra were calculated by sweeping the energy loss of the incident electron beam. The peaks in these spectra were fitted with Lorentzian profiles to extract the central energies of the dipole LSPR modes for each PA. The loss function maps were obtained by scanning the electron beam position over a defined area around the PA at the peak energy loss of the dipole LSPR. These calculations were carried out for a set of PAs (diamond, rod, and dumbbell geometries) with identical length $L=300$~nm, radius $R=50$~nm, and thickness of 30~nm, but varying widths (listed in Table~\ref{tab1}).

The dipole resonance energies of individual PAs can be tuned by modifying their lengths. A resonance energy of 1.2~eV was found to be achievable for all PA types with reasonable length adjustments. To reach this energy, the lengths of the individual PA types were varied from the nominal value of 300~nm to the values listed in the lower part of Table~\ref{tab1}. Note that the widths of the diamond PA and the tapered diamond PA were adjusted accordingly to preserve their geometrical aspect ratios, while the widths of the thinner PAs (rod and dumbbell types) remained unchanged. Similarly, we obtained the dimensions of the PA with the resonance energy fixed at 0.8~eV and 1.7~eV. Due to the size restrictions of PAs with the resonance energy of 1.7~eV, we have adjusted their radius $R$ to 20~nm.

Next, the excitation was changed from the electron beam to a plane electromagnetic wave polarized along the horizontal axis of the PAs. Scattering and absorption spectra were calculated from the induced surface charges and currents by varying the energy of the incident plane wave. For the fixed peak energies, the induced electric fields were also computed and quantified.


\section{Effect of the substrate}

Throughout the main paper, the substrate is omitted from the simulations. In this section, we discuss its effect on the loss function spectra. In MNPBEM, the silicon nitride (SiN) membrane used in the experiments was modeled as two parallel plates separated by a distance corresponding to the membrane thickness (30~nm), on top of which the particle was placed, as shown in Fig.~\ref{fig1SI}(c). The dielectric environment of the SiN membrane can be approximated by a dielectric constant set equal to 4 in the considered spectral region~\cite{Schmidt2014}.
 
\begin{figure*}
\includegraphics[width=1.5\columnwidth]{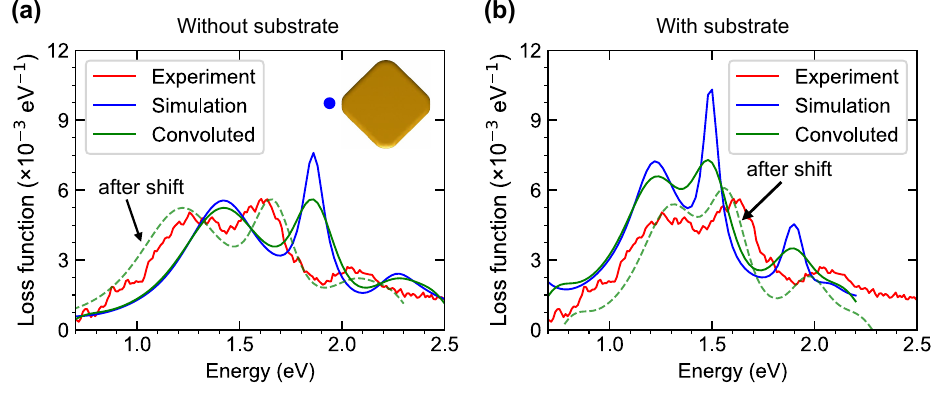}
\centering
\caption{Comparison of experimental loss function spectra (red lines) with simulated spectra (blue lines) for two models: (a) PA without the substrate and (b) PA on a substrate. The electron beam position is indicated in the inset (blue dot). The calculated spectra were convolved with a Gaussian function (full width at half maximum of 0.15~eV; green line) and manually shifted in energy for comparison (green dashed line).}

\label{fig2SI}

\end{figure*}

For comparison, the loss function spectra were calculated for a diamond-shaped PA with a diagonal length of 300~nm, first without the substrate [Fig.~\ref{fig2SI}(a)] and then with the substrate included [Fig.~\ref{fig2SI}(b)]. The electron beam was positioned 20~nm from the left edge of the antenna (indicated by the blue dot), and its incident energy was set to 120~keV. To account for instrumental broadening, mainly due to the finite width of the zero-loss peak, the calculated spectra were convolved with a Gaussian function of 0.15~eV full width at half maximum (FWHM), shown as solid green lines.

Let us first consider the case without the substrate [Fig.~\ref{fig2SI}(a)]. A comparison of the calculated (blue) and experimental (red) spectra shows the following: (1) The calculated spectrum is blue-shifted by approximately 0.15~eV with respect to the experiment. (2) The first two LSPR peaks are slightly more separated in the calculated spectrum (0.44~eV) than in the experimental one (0.36~eV). The convolved spectrum (green line) reproduces the experimental spectrum reasonably well, except for a slight energy offset, which can be primarily attributed to the omission of the substrate. To facilitate comparison, a constant energy shift of 0.2~eV was applied to the convolved spectrum; the resulting spectrum (green dashed line) closely matches the experimental result.

A similar analysis can be performed for the case with the substrate included [Fig.~\ref{fig2SI}(b)]. The following differences can be observed: (1) The calculated spectrum is red-shifted by approximately 0.1~eV relative to the experiment. This energy shift may arise from by differences between the real and model dimensions of the PA or from variations in the real and model dielectric functions. (2) The peaks in the calculated spectrum are slightly less separated (0.28~eV) than in the experimental one (0.36~eV). (3) The calculated spectrum exhibits a higher background signal, which in the experiment is partially removed by subtracting the membrane contribution. As before, the calculated spectrum was convolved with the Gaussian function (green line) and shifted in both energy (by 0.1~eV) and intensity (by $2\times10^{-3}$~eV$^{-1}$) to enable direct comparison with the experiment (green dashed line).

In summary, the comparison of experimental and simulated loss function spectra demonstrates that the overall spectral features are well reproduced even when the substrate is omitted from the simulations, with only a minor shift in the peak energies. Including the substrate introduces additional numerical complexity and potential inconsistencies in the calculations, particularly due to the thin dielectric layer and its interface treatment. Therefore, to ensure numerical stability and focus on the plasmonic behavior of the antennas, the substrate was omitted from all simulations presented in the main paper.

\section{Generalized harmonic oscillator model}
Kats {\it et al.}~\cite{Kats:11} proposed a plasmon model in which a body of electrons with the charge $q$, the mass $m$, and the displacement $x(t)$ is driven by the harmonic external electric field $E_0\exp( i\omega t)$ and experiences the restoring force $m\omega_0^2$, the internal damping $\Gamma_a \mathrm{d}x/\mathrm{d}t$, and the radiative reaction force $\Gamma_s \mathrm{d}^3x/\mathrm{d}t^3$ with $\Gamma_s=q^2/6\pi\varepsilon_0 c^3$. The equation of the motion thus reads
\begin{equation}
m\frac{\mathrm{d}^2x}{\mathrm{d}t^2} + \Gamma_a \frac{\mathrm{d}x}{\mathrm{d}t} + m\omega_0^2 x = qE_0 \exp (i\omega t) + \Gamma_s \frac{\mathrm{d}^3x}{\mathrm{d}t^3}.
\end{equation} 

The amplitude of the displacement $x_0$ follows
\begin{equation}
|x_0(\omega)|^2=\frac{q^2}{m^2}\frac{E_0^2}{(\omega_0^2-\omega^2)^2+(\omega^2/m^2)(\Gamma_a+\omega^2\Gamma_s)^2}.
\end{equation} 

Finally, considering the usual expression for the plane wave intensity
\begin{equation}
I_0=\frac{E_0^2}{2\mu_0 c},
\end{equation}
we obtain the absorption cross-section
\begin{equation}
C_\mathrm{abs}(\omega)=2\mu_0 c \omega^2 \Gamma_a \frac{|x_0(\omega)|^2}{E_0^2},
\label{eqabs}
\end{equation}
and the scattering cross-section
\begin{equation}
C_\mathrm{scat}(\omega)=2\mu_0 c \omega^4 \Gamma_s \frac{|x_0(\omega)|^2}{E_0^2}.
\label{eqscat}
\end{equation}
Both cross-sections contain four free parameters ($\omega_0$, $q$, $m$, $\Gamma_a$) that can be determined by fitting the functions (Eqs.~\ref{eqabs} and \ref{eqscat}) to the simulated data.

We have slightly adapted the model to our purposes. First, we have noticed that the model is overparameterized. The three parameters, $q$, $m$, and $\Gamma_a$, are included in Eqs.~\ref{eqabs} and \ref{eqscat} only in two expressions, $q^2/m$ and $\Gamma_a/m$. It is thus impossible to determine all three parameters. Instead, we determine the total mass of the electrons as $m=Nm^*m_0$, where the number of electrons $N=q/e$ ($e$ is the elementary charge), $m_0$ is the electron mass, and for the effective mass of gold $m^*$ we take the value 1.35 reported in Ref.~\cite{Kats:11}. 

Further, we included a phenomenological amplitude factor $C_0$, which is needed for a reliable reproduction of the amplitude of both cross-sections. The need for this factor can stem primarily from the overly simple oscillator model that neglects, among other factors, the spatial distribution of the induced electromagnetic field. Further need for $C_0$ might arise from the limited accuracy of the numerical simulations yielding the data to which the model is fitted. The adapted spectral dependences of the optical cross-sections read
\begin{equation}
C_\mathrm{abs}(\omega)=2 C_0 \mu_0 c \omega^2 \Gamma_a \frac{|x_0(\omega)|^2}{E_0^2},
\label{eqabsm}
\end{equation}
\begin{equation}
C_\mathrm{scat}(\omega)=2 C_0 \mu_0 c \omega^4 \Gamma_s \frac{|x_0(\omega)|^2}{E_0^2}.
\label{eqscatm}
\end{equation}
with free parameters $C_0$, $\omega_0$, $q$, and $\Gamma_a$.

\begin{figure*}
\begin{center}
\includegraphics[width=1.5\columnwidth]{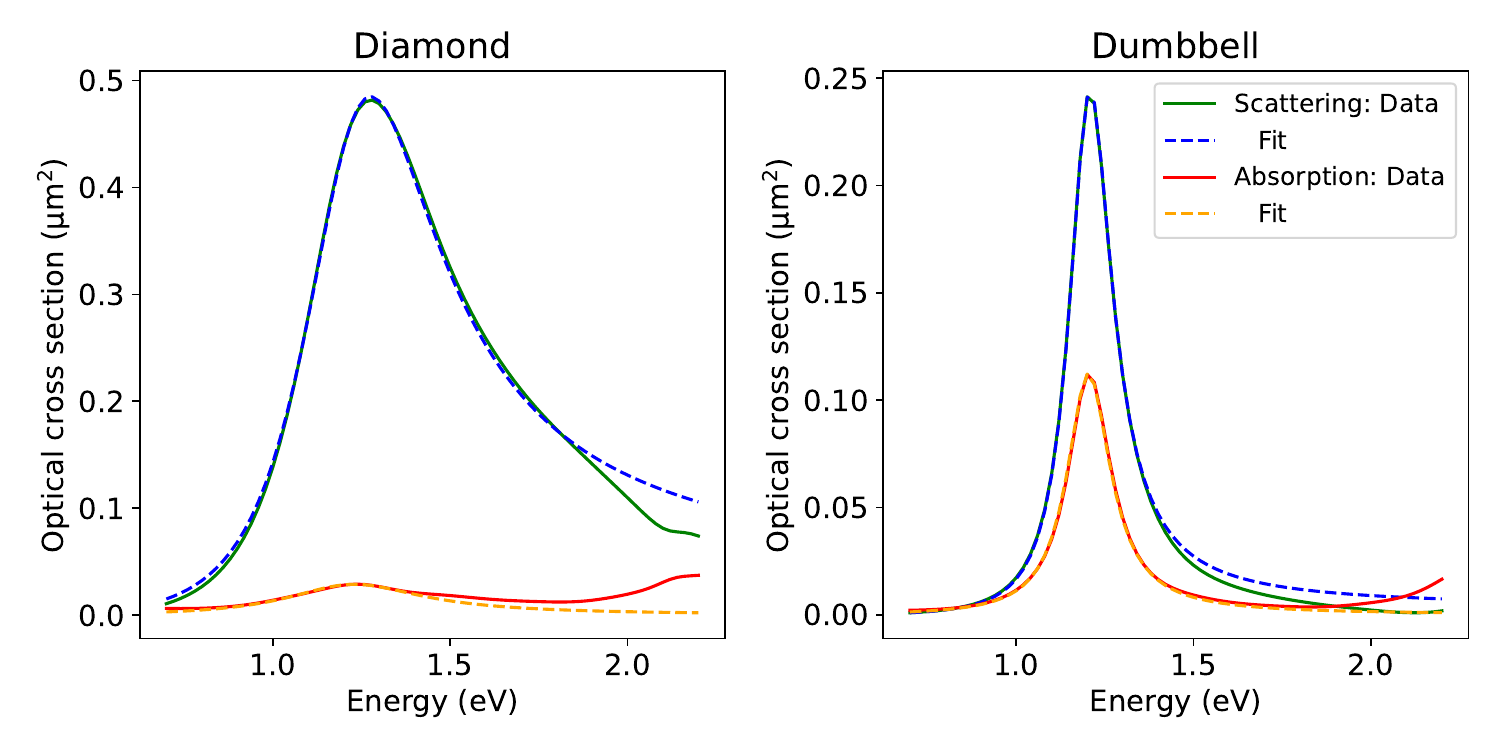}
\caption{
Calculated scattering cross-section (green solid line) fitted with the model of Eq.~\ref{eqscatm} (blue dashed line) and calculated absorption cross-section (red solid line) fitted with the model of Eq.~\ref{eqabsm} (orange dashed line) for the PAs with the dipole LSPR energy of 1.2~eV: (left) A diamond with a width of 385~nm. (right) A dumbbell with a width of 40~nm.  
\label{figSI3}
}
\end{center}
\end{figure*}

The models of the absorption and scattering cross-sections (Eqs.~\ref{eqabsm} and \ref{eqscatm}) were fitted to the simulated cross-sections shown in Fig.~4(a,b) of the main paper and similar data obtained for other sets of PAs. The accuracy of the models is very good, as demonstrated in Fig.~\ref{figSI3} for the instance of two PAs with the dipole LSPR energy of 1.2 eV. This figure clearly shows that the optical response in the broad spectral range can be assigned to a single dipole LSPR, confirming our assignment of the higher-order peak in the EELS spectrum [Fig.~2(b,d) in the main paper] to a dark mode. Minor deviations above the energy of 1.8~eV (a decrease in the calculated scattering cross-section and an increase in the calculated absorption cross-section compared to the model quantities) can be attributed to the onset of the interband absorption of gold, which is not included in the oscillator model.

\bibliographystyle{IEEEtran}


\begin{thebibliography}{10}
\providecommand{\url}[1]{#1}
\csname url@samestyle\endcsname
\providecommand{\newblock}{\relax}
\providecommand{\bibinfo}[2]{#2}
\providecommand{\BIBentrySTDinterwordspacing}{\spaceskip=0pt\relax}
\providecommand{\BIBentryALTinterwordstretchfactor}{4}
\providecommand{\BIBentryALTinterwordspacing}{\spaceskip=\fontdimen2\font plus
\BIBentryALTinterwordstretchfactor\fontdimen3\font minus
  \fontdimen4\font\relax}
\providecommand{\BIBforeignlanguage}[2]{{%
\expandafter\ifx\csname l@#1\endcsname\relax
\typeout{** WARNING: IEEEtran.bst: No hyphenation pattern has been}%
\typeout{** loaded for the language `#1'. Using the pattern for}%
\typeout{** the default language instead.}%
\else
\language=\csname l@#1\endcsname
\fi
#2}}
\providecommand{\BIBdecl}{\relax}
\BIBdecl

\bibitem{Novotny2011}
L.~Novotny and N.~van Hulst, ``Antennas for light,'' \emph{Nat. Photonics},
  vol.~5, no.~2, pp. 83--90, 2011.

\bibitem{Kinkhabwala2009}
A.~Kinkhabwala, Z.~Yu, S.~Fan, Y.~Avlasevich, K.~M{\"u}llen, and W.~E. Moerner,
  ``Large single-molecule fluorescence enhancements produced by a bowtie
  nanoantenna,'' \emph{Nat. Photonics}, vol.~3, p. 654, 2009.

\bibitem{Pfeiffer2010}
M.~Pfeiffer, K.~Lindfors, C.~Wolpert, P.~Atkinson, M.~Benyoucef, A.~Rastelli,
  O.~G. Schmidt, H.~Giessen, and M.~Lippitz, ``Enhancing the optical excitation
  efficiency of a single self-assembled quantum dot with a plasmonic
  nanoantenna,'' \emph{Nano Lett.}, vol.~10, no.~11, pp. 4555--4558, 2010.

\bibitem{Wang2020}
X.~Wang, S.-C. Huang, S.~Hu, S.~Yan, and B.~Ren, ``Fundamental understanding
  and applications of plasmon-enhanced {R}aman spectroscopy,'' \emph{Nat. Rev.
  Phys.}, vol.~2, no.~5, pp. 253--271, 2020.

\bibitem{Han2022}
X.~X. Han, R.~S. Rodriguez, C.~L. Haynes, Y.~Ozaki, and B.~Zhao,
  ``Surface-enhanced {R}aman spectroscopy,'' \emph{Nat. Rev. Methods Primers},
  vol.~1, no.~1, p.~87, 2022.

\bibitem{https://doi.org/10.1002/smtd.202100376}
L.~Tesi, D.~Bloos, M.~Hrtoň, A.~Beneš, M.~Hentschel, M.~Kern, A.~Leavesley,
  R.~Hillenbrand, V.~Křápek, T.~Šikola, and J.~van Slageren, ``Plasmonic
  metasurface resonators to enhance terahertz magnetic fields for
  high-frequency electron paramagnetic resonance,'' \emph{Small Methods},
  vol.~5, no.~9, p. 2100376, 2021.

\bibitem{BORISKINA2013375}
S.~V. Boriskina, H.~Ghasemi, and G.~Chen, ``Plasmonic materials for energy:
  From physics to applications,'' \emph{Mater. Today}, vol.~16, no.~10, pp.
  375--386, 2013.

\bibitem{doi:10.1021/nl104352j}
N.~T. Fofang, N.~K. Grady, Z.~Fan, A.~O. Govorov, and N.~J. Halas, ``Plexciton
  dynamics: Exciton-plasmon coupling in a j-aggregate-{A}u nanoshell complex
  provides a mechanism for nonlinearity,'' \emph{Nano Lett.}, vol.~11, no.~4,
  pp. 1556--1560, 2011.

\bibitem{Bitton2020}
O.~Bitton, S.~N. Gupta, L.~Houben, M.~Kvapil, V.~K{\v{r}}{\'a}pek,
  T.~{\v{S}}ikola, and G.~Haran, ``Vacuum {R}abi splitting of a dark plasmonic
  cavity mode revealed by fast electrons,'' \emph{Nat. Commun.}, vol.~11,
  no.~1, p. 487, 2020.

\bibitem{annurev:/content/journals/10.1146/annurev.physchem.58.032806.104607}
K.~A. Willets and R.~P. Van~Duyne, ``Localized surface plasmon resonance
  spectroscopy and sensing,'' \emph{Annu. Rev. Phys. Chem}, vol.~58, pp.
  267--297, 2007.

\bibitem{Riley2023}
J.~A. Riley, M.~Horák, V.~Křápek, N.~Healy, and V.~Pacheco-Peña,
  ``Plasmonic sensing using {B}abinet’s principle,'' \emph{Nanophotonics},
  vol.~12, no.~20, pp. 3895--3909, 2023.

\bibitem{Mejia-Salazar2018}
J.~R. Mej{\'i}a-Salazar and O.~N. Oliveira~Jr., ``Plasmonic biosensing,''
  \emph{Chem. Rev.}, vol. 118, no.~20, pp. 10\,617--10\,625, 2018.

\bibitem{PhysRevLett.95.095504}
N.~Engheta, A.~Salandrino, and A.~Al\`u, ``Circuit elements at optical
  frequencies: Nanoinductors, nanocapacitors, and nanoresistors,'' \emph{Phys.
  Rev. Lett.}, vol.~95, p. 095504, 2005.

\bibitem{Zhu:14}
D.~Zhu, M.~Bosman, and J.~K.~W. Yang, ``A circuit model for plasmonic
  resonators,'' \emph{Opt. Express}, vol.~22, no.~8, pp. 9809--9819, 2014.

\bibitem{Benz:15}
F.~Benz, B.~de~Nijs, C.~Tserkezis, R.~Chikkaraddy, D.~O. Sigle, L.~Pukenas,
  S.~D. Evans, J.~Aizpurua, and J.~J. Baumberg, ``Generalized circuit model for
  coupled plasmonic systems,'' \emph{Opt. Express}, vol.~23, no.~26, pp.
  33\,255--33\,269, 2015.

\bibitem{Hughes2016}
T.~W. Hughes and S.~Fan, ``Plasmonic circuit theory for multiresonant light
  funneling to a single spatial hot spot,'' \emph{Nano Lett.}, vol.~16, no.~9,
  pp. 5764--5769, 2016.

\bibitem{doi:10.1021/nl303377g}
A.~I. Fernández-Domínguez, Y.~Luo, A.~Wiener, J.~B. Pendry, and S.~A. Maier,
  ``Theory of three-dimensional nanocrescent light harvesters,'' \emph{Nano
  Lett.}, vol.~12, no.~11, pp. 5946--5953, 2012.

\bibitem{doi:10.1021/acs.jpcc.2c04828}
R.~A. Alves, V.~Pacheco-Pena, and M.~Navarro-Cía, ``Madelung formalism for
  electron spill-out in nonlocal nanoplasmonics,'' \emph{J. Phys. Chem. C},
  vol. 126, no.~34, pp. 14\,758--14\,765, 2022.

\bibitem{doi:10.1126/science.1089171}
E.~Prodan, C.~Radloff, N.~J. Halas, and P.~Nordlander, ``A hybridization model
  for the plasmon response of complex nanostructures,'' \emph{Science}, vol.
  302, no. 5644, pp. 419--422, 2003.

\bibitem{krapek_independent}
V.~Křápek, A.~Konečná, M.~Horák, F.~Ligmajer, M.~Stöger-Pollach,
  M.~Hrtoň, J.~Babocký, and T.~Šikola, ``Independent engineering of
  individual plasmon modes in plasmonic dimers with conductive and capacitive
  coupling,'' \emph{Nanophotonics}, vol.~9, no.~3, pp. 623--632, 2020.

\bibitem{PhysRevB.76.033407}
T.~Zentgraf, T.~P. Meyrath, A.~Seidel, S.~Kaiser, H.~Giessen, C.~Rockstuhl, and
  F.~Lederer, ``Babinet's principle for optical frequency metamaterials and
  nanoantennas,'' \emph{Phys. Rev. B}, vol.~76, p. 033407, 2007.

\bibitem{doi:10.1021/nl402269h}
M.~Hentschel, T.~Weiss, S.~Bagheri, and H.~Giessen, ``Babinet to the half:
  Coupling of solid and inverse plasmonic structures,'' \emph{Nano Lett.},
  vol.~13, no.~9, pp. 4428--4433, 2013.

\bibitem{Horak2019}
M.~Hor{\'a}k, V.~K{\v{r}}{\'a}pek, M.~Hrto\v{n}, A.~Kone\v{c}n{\'a},
  F.~Ligmajer, M.~St{\"o}ger-Pollach, T.~\v{S}amo\v{r}il, A.~Pat{\'a}k,
  Z.~{\'E}des, O.~Metelka, J.~Babock{\'y}, and T.~\v{S}ikola, ``Limits of
  {B}abinet's principle for solid and hollow plasmonic antennas,'' \emph{Sci.
  Rep.}, vol.~9, no.~1, p. 4004, 2019.

\bibitem{10.1063/5.0065724}
J.~D. Ortiz, J.~P. del Risco, J.~D. Baena, and R.~Marqués, ``{Extension of
  Babinet's principle for plasmonic metasurfaces},'' \emph{Appl. Phys. Lett.},
  vol. 119, no.~16, p. 161103, 2021.

\bibitem{10.1063/1.372414}
A.~J. Babadjanyan, N.~L. Margaryan, and K.~V. Nerkararyan, ``Superfocusing of
  surface polaritons in the conical structure,'' \emph{J. Appl. Phys.},
  vol.~87, no.~8, pp. 3785--3788, 2000.

\bibitem{PhysRevLett.93.137404}
M.~I. Stockman, ``Nanofocusing of optical energy in tapered plasmonic
  waveguides,'' \emph{Phys. Rev. Lett.}, vol.~93, p. 137404, 2004.

\bibitem{doi:10.1021/nl071340m}
C.~Ropers, C.~C. Neacsu, T.~Elsaesser, M.~Albrecht, M.~B. Raschke, and
  C.~Lienau, ``Grating-coupling of surface plasmons onto metallic tips: A
  nanoconfined light source,'' \emph{Nano Lett.}, vol.~7, no.~9, pp.
  2784--2788, 2007.

\bibitem{PhysRevLett.105.116804}
A.~R. Davoyan, I.~V. Shadrivov, A.~A. Zharov, D.~K. Gramotnev, and Y.~S.
  Kivshar, ``Nonlinear nanofocusing in tapered plasmonic waveguides,''
  \emph{Phys. Rev. Lett.}, vol. 105, p. 116804, 2010.

\bibitem{Schnell2011}
M.~Schnell, P.~Alonso-Gonz{\'a}lez, L.~Arzubiaga, F.~Casanova, L.~E. Hueso,
  A.~Chuvilin, and R.~Hillenbrand, ``Nanofocusing of mid-infrared energy with
  tapered transmission lines,'' \emph{Nat. Photon.}, vol.~5, no.~5, pp.
  283--287, 2011.

\bibitem{Choo2012}
H.~Choo, M.-K. Kim, M.~Staffaroni, T.~J. Seok, J.~Bokor, S.~Cabrini, P.~J.
  Schuck, M.~C. Wu, and E.~Yablonovitch, ``Nanofocusing in a
  metal--insulator--metal gap plasmon waveguide with a three-dimensional linear
  taper,'' \emph{Nat. Photon.}, vol.~6, no.~12, pp. 838--844, 2012.

\bibitem{Gramotnev2014}
D.~K. Gramotnev and S.~I. Bozhevolnyi, ``Nanofocusing of electromagnetic
  radiation,'' \emph{Nat. Photon.}, vol.~8, no.~1, pp. 13--22, 2014.

\bibitem{9172141}
F.~Lu, W.~Zhang, M.~Liu, L.~Zhang, and T.~Mei, ``Tip-based plasmonic
  nanofocusing: Vector field engineering and background elimination,''
  \emph{IEEE J. Sel. Top. Quantum Electron.}, vol.~27, no.~1, pp. 1--12, 2021.

\bibitem{Garcia-Etxarri:12}
A.~Garc\'{i}a-Etxarri, P.~Apell, M.~K\"{a}ll, and J.~Aizpurua, ``A combination
  of concave/convex surfaces for field-enhancement optimization: the indented
  nanocone,'' \emph{Opt. Express}, vol.~20, no.~23, pp. 25\,201--25\,212, 2012.

\bibitem{Shi:17}
L.~Shi, B.~Iwan, R.~Nicolas, Q.~Ripault, J.~R.~C. Andrade, S.~Han, H.~Kim,
  W.~Boutu, D.~Franz, T.~Heidenblut, C.~Reinhardt, B.~Bastiaens, T.~Nagy,
  I.~Babushkin, U.~Morgner, S.-W. Kim, G.~Steinmeyer, H.~Merdji, and
  M.~Kovacev, ``Self-optimization of plasmonic nanoantennas in strong
  femtosecond fields,'' \emph{Optica}, vol.~4, no.~9, pp. 1038--1043, 2017.

\bibitem{Jie_Yang}
J.~Yang, F.~Kong, K.~Li, and J.~Zhao, ``Optimizing the bowtie nano-antenna for
  enhanced {P}urcell factor and electric field,'' \emph{Prog. Electromagn. Res.
  Lett.}, vol.~44, pp. 93--99, 2014.

\bibitem{Rogobete:07}
L.~Rogobete, F.~Kaminski, M.~Agio, and V.~Sandoghdar, ``Design of plasmonic
  nanoantennae for enhancing spontaneous emission,'' \emph{Opt. Lett.},
  vol.~32, no.~12, pp. 1623--1625, 2007.

\bibitem{Urbieta2018}
M.~Urbieta, M.~Barbry, Y.~Zhang, P.~Koval, D.~S{\'a}nchez-Portal, N.~Zabala,
  and J.~Aizpurua, ``Atomic-scale lightning rod effect in plasmonic
  picocavities: A classical view to a quantum effect,'' \emph{ACS Nano},
  vol.~12, no.~1, pp. 585--595, 2018.

\bibitem{krapek_PLRE}
V.~Křápek, R.~Řepa, M.~Foltýn, T.~Šikola, and M.~Horák, ``Plasmonic
  lightning-rod effect,'' arXiv:2407.09454, 2024.

\bibitem{Biagioni_2012}
P.~Biagioni, J.-S. Huang, and B.~Hecht, ``Nanoantennas for visible and infrared
  radiation,'' \emph{Rep. Prog. Phys.}, vol.~75, no.~2, p. 024402, 2012.

\bibitem{10.1063/1.1602956}
K.~B. Crozier, A.~Sundaramurthy, G.~S. Kino, and C.~F. Quate, ``Optical
  antennas: Resonators for local field enhancement,'' \emph{J. Appl. Phys.},
  vol.~94, no.~7, pp. 4632--4642, 2003.

\bibitem{Chen:17}
C.~Chen, G.~Wang, L.~Peng, and K.~Zhang, ``Highly improved, non-localized field
  enhancement enabled by hybrid plasmon of crescent resonator/graphene in
  infrared wavelength,'' \emph{Opt. Express}, vol.~25, no.~19, pp.
  23\,302--23\,311, Sep 2017.

\bibitem{Li:25}
T.~Li, A.~Schirato, T.~Suwabe, R.~P. Zaccaria, P.~Verma, and T.~Umakoshi,
  ``Comparison of near-field light intensities: plasmon nanofocusing versus
  localized plasmon resonance,'' \emph{Opt. Express}, vol.~33, no.~13, pp.
  26\,930--26\,938, 2025.

\bibitem{Sederberg:11}
S.~Sederberg and A.~Y. Elezzabi, ``Nanoscale plasmonic contour bowtie antenna
  operating in the mid-infrared,'' \emph{Opt. Express}, vol.~19, no.~16, pp.
  15\,532--15\,537, 2011.

\bibitem{doi:10.1021/jp408610q}
L.-W. Nien, S.-C. Lin, B.-K. Chao, M.-J. Chen, J.-H. Li, and C.-H. Hsueh,
  ``Giant electric field enhancement and localized surface plasmon resonance by
  optimizing contour bowtie nanoantennas,'' \emph{J. Phys. Chem. C}, vol. 117,
  no.~47, pp. 25\,004--25\,011, 2013.

\bibitem{Sederberg:11b}
S.~Sederberg and A.~Elezzabi, ``Sierpi\'{n}ski fractal plasmonic antenna: a
  fractal abstraction of the plasmonic bowtie antenna,'' \emph{Opt. Express},
  vol.~19, no.~11, pp. 10\,456--10\,461, 2011.

\bibitem{Horak2018}
M.~Hor{\'a}k, K.~Bukvi\v{s}ov{\'a}, V.~{\v{S}}varc, J.~Jaskowiec,
  V.~K{\v{r}}{\'a}pek, and T.~\v{S}ikola, ``Comparative study of plasmonic
  antennas fabricated by electron beam and focused ion beam lithography,''
  \emph{Sci. Rep.}, vol.~8, no.~1, p. 9640, 2018.

\bibitem{doi:10.1021/acsomega.4c06598}
M.~Foltýn, M.~Patočka, R.~Řepa, T.~Šikola, and M.~Horák, ``Influence of
  deposition parameters on the plasmonic properties of gold nanoantennas
  fabricated by focused ion beam lithography,'' \emph{ACS Omega}, vol.~9,
  no.~35, pp. 37\,408--37\,416, 2024.

\bibitem{HOHENESTER2012370}
U.~Hohenester and A.~Trügler, ``{MNPBEM} -- a {M}atlab toolbox for the
  simulation of plasmonic nanoparticles,'' \emph{Comput. Phys. Commun.}, vol.
  183, no.~2, pp. 370--381, 2012.

\bibitem{HOHENESTER20141177}
U.~Hohenester, ``Simulating electron energy loss spectroscopy with the {MNPBEM}
  toolbox,'' \emph{Computer Physics Communications}, vol. 185, no.~3, pp.
  1177--1187, 2014.

\bibitem{WAXENEGGER2015138}
J.~Waxenegger, A.~Trügler, and U.~Hohenester, ``Plasmonics simulations with
  the {MNPBEM} toolbox: {C}onsideration of substrates and layer structures,''
  \emph{Comput. Phys. Commun.}, vol. 193, pp. 138--150, 2015.

\bibitem{PhysRevB.65.115418}
F.~J. Garc\'{\i}a~de Abajo and A.~Howie, ``Retarded field calculation of
  electron energy loss in inhomogeneous dielectrics,'' \emph{Phys. Rev. B},
  vol.~65, p. 115418, 2002.

\bibitem{PhysRevB.6.4370}
P.~B. Johnson and R.~W. Christy, ``Optical constants of the noble metals,''
  \emph{Phys. Rev. B}, vol.~6, pp. 4370--4379, 1972.

\bibitem{Kejik:20}
L.~Kej\'{i}k, M.~Hor\'{a}k, T.~\v{S}ikola, and V.~K\v{r}\'{a}pek, ``Structural
  and optical properties of monocrystalline and polycrystalline gold plasmonic
  nanorods,'' \emph{Opt. Express}, vol.~28, no.~23, pp. 34\,960--34\,972, 2020.

\bibitem{RevModPhys.82.209}
F.~J. Garc\'{\i}a~de Abajo, ``Optical excitations in electron microscopy,''
  \emph{Rev. Mod. Phys.}, vol.~82, pp. 209--275, 2010.

\bibitem{Kats:11}
M.~A. Kats, N.~Yu, P.~Genevet, Z.~Gaburro, and F.~Capasso, ``Effect of
  radiation damping on the spectral response of plasmonic components,''
  \emph{Opt. Express}, vol.~19, no.~22, pp. 21\,748--21\,753, 2011.

\bibitem{https://doi.org/10.1002/adom.202001520}
B.~Wang, P.~Yu, W.~Wang, X.~Zhang, H.-C. Kuo, H.~Xu, and Z.~M. Wang, ``High-{Q}
  plasmonic resonances: Fundamentals and applications,'' \emph{Adv. Opt.
  Mater.}, vol.~9, no.~7, p. 2001520, 2021.

\bibitem{PhysRevLett.88.077402}
C.~S\"onnichsen, T.~Franzl, T.~Wilk, G.~von Plessen, J.~Feldmann, O.~Wilson,
  and P.~Mulvaney, ``Drastic reduction of plasmon damping in gold nanorods,''
  \emph{Phys. Rev. Lett.}, vol.~88, p. 077402, Jan 2002.

\bibitem{KOLWAS201345}
K.~Kolwas and A.~Derkachova, ``Damping rates of surface plasmons for particles
  of size from nano- to micrometers; reduction of the nonradiative decay,''
  \emph{J. Quant. Spectrosc. Radiat. Transf.}, vol. 114, pp. 45--55, 2013.

\bibitem{10.1063/1.5117230}
T.~Devkota, B.~S. Brown, G.~Beane, K.~Yu, and G.~V. Hartland, ``Making waves:
  Radiation damping in metallic nanostructures,'' \emph{J. Chem. Phys.}, vol.
  151, no.~8, p. 080901, 2019.

\bibitem{HU2022100166}
H.~Hu, W.~Tao, F.~Laible, T.~Maurer, P.-M. Adam, A.~Horneber, and M.~Fleischer,
  ``Spectral exploration of asymmetric bowtie nanoantennas,'' \emph{Micro and
  Nano Engineering}, vol.~17, p. 100166, 2022.

\bibitem{10.1063/1.2963699}
D.~K. Gramotnev, M.~W. Vogel, and M.~I. Stockman, ``Optimized nonadiabatic
  nanofocusing of plasmons by tapered metal rods,'' \emph{J. Appl. Phys.}, vol.
  104, no.~3, p. 034311, 2008.

\bibitem{doi:10.1126/sciadv.aba4179}
T.~Umakoshi, M.~Tanaka, Y.~Saito, and P.~Verma, ``White nanolight source for
  optical nanoimaging,'' \emph{Science Advances}, vol.~6, no.~23, p. eaba4179,
  2020.

\bibitem{Ma2021}
X.~Ma, Q.~Liu, N.~Yu, D.~Xu, S.~Kim, Z.~Liu, K.~Jiang, B.~M. Wong, R.~Yan, and
  M.~Liu, ``6{\thinspace}nm super-resolution optical transmission and
  scattering spectroscopic imaging of carbon nanotubes using a nanometer-scale
  white light source,'' \emph{Nat. Commun.}, vol.~12, no.~1, p. 6868, 2021.

\bibitem{doi:10.1021/acs.jpcc.0c11541}
K.~Taguchi, T.~Umakoshi, S.~Inoue, and P.~Verma, ``Broadband plasmon
  nanofocusing: Comprehensive study of broadband nanoscale light source,''
  \emph{J. Phys. Chem. C}, vol. 125, no.~11, pp. 6378--6386, 2021.

\bibitem{Schmidt2014}
F.~P. Schmidt, H.~Ditlbacher, F.~Hofer, J.~R. Krenn, and U.~Hohenester,
  ``Morphing a plasmonic nanodisk into a nanotriangle,'' \emph{Nano Lett.},
  vol.~14, no.~8, pp. 4810--4815, 2014.

\end{thebibliography}

\end{document}